\documentclass[12pt]{article}
\usepackage{axodraw}
\input epsf.tex

\newcommand{\beq}{\begin{equation}}
\newcommand{\eeq}{\end{equation}}
\newcommand{\beqa}{\begin{eqnarray}}
\newcommand{\eeqa}{\end{eqnarray}}
\newcommand{\beqar}{\begin{eqnarray*}}
\newcommand{\eeqar}{\end{eqnarray*}}

\newcommand{\be}{\beta}

\newcommand{\eps}{\epsilon}
\newcommand{\ga}{\gamma}
\newcommand{\Ga}{\Gamma}
\newcommand{\ka}{\kappa}
\newcommand{\inn}{\!\cdot\!}
\newcommand{\h}{\eta}

\newcommand{\la}{\lambda}

\renewcommand{\t}{\theta}

\newcommand{\eg}{{\it e.g.,}\ }
\newcommand{\ie}{{\it i.e.,}\ }
\newcommand{\labell}[1]{\label{#1}} 
\newcommand{\reef}[1]{(\ref{#1})}
\newcommand\prt{\partial}
\newcommand\veps{\varepsilon}

\newcommand\cF{{\cal F}}
\newcommand\cA{{\cal A}}

\newcommand\cC{{\cal C}}

\newcommand\cI{{\cal I}}

\newcommand\cP{{\cal P}}

\newcommand\bz{\bar{z}}

\newcommand\tK{{\widetilde K}}

\newcommand\Tr{{\rm Tr}}

\begin{document}

\vspace*{1cm}

\begin{center}
{\bf \Large  On RR couplings on D-branes at order $O(\alpha'^2)$}

\vspace*{1cm}

{Mohammad R. Garousi and Mozhgan Mir}\\
\vspace*{1cm}
{ Department of Physics, Ferdowsi University of Mashhad,\\ P.O. Box 1436, Mashhad, Iran}
\\
\vspace{2cm}

\end{center}

\begin{abstract}
\baselineskip=18pt

Recently,  it has been found that there are  couplings of the RR field strength $F^{(p)}$ and the  B-field strength $H$   on the world volume of D$_p$-branes at order ${\cal O}(\alpha'^2)$.  These couplings which have both world-volume and transverse indices, are  invariant under the linear T-duality  transformations.  Consistency with the nonlinear T-duality  indicates  that the RR field strength $F^{(p)}$ in these couplings should be replaced by ${\cal F}^{(p)}=d{\cal C}^{(p-1)}$ where ${\cal C}=e^{B}C$. This replacement, however, produces some non-gauge invariant terms. On the other hand,  the nonlinear terms are invariant under the linear T-duality transformations at the level of two B-fields. This  allows one to remove some of the nonlinear terms in ${\cal F}^{(p)}$. We fix this  by comparing the nonlinear couplings with the S-matrix element of one RR and two NSNS vertex operators. Our results indicate that  in the expansion of  ${\cal F}^{(p)}$ one should keep only the B-field gauge invariant terms, \eg $B\wedge dC^{(p-3)}$ where both indices of B-field lie along the brane. Moreover, in this case one should replace $B$ with $B+2\pi\alpha'f$ to have the  $B$-field gauge invariance.

\end{abstract}

\vfill
\setcounter{page}{0}
\setcounter{footnote}{0}
\newpage


\section{Introduction } \label{intro}

The dynamics of the D-branes of type II superstring theories is well-approximated by the effective world-volume theory which is  sum of   Dirac-Born-Infeld (DBI) and Chern-Simons (CS) actions. 
The DBI action  which describes the dynamics of the branes in the presence of the NSNS background fields at order ${\cal O}(\alpha'^0)$ is given by \cite{Leigh:1989jq,Bachas:1995kx}
\beqa
S_{DBI}&=&-T_p\int d^{p+1}x\,e^{-\phi}\sqrt{-\det\left(G_{ab}+B_{ab}\right)}\labell{DBI}
\eeqa
where $G_{ab}$ and $B_{ab}$ are  the pulled back of the bulk fields $G_{\mu\nu}$ and $B_{\mu\nu}$ onto the world-volume of D-brane\footnote{Our index conversion is that the Greek letters  $(\mu,\nu,\cdots)$ are  the indices of the space-time coordinates, the Latin letters $(a,d,c,\cdots)$ are the world-volume indices and the letters $(i,j,k,\cdots)$ are the normal bundle indices.}. The abelian gauge field can be added to the action as $B_{ab}\rightarrow B_{ab}+2\pi\alpha'f_{ab}$. The curvature corrections to this action has been found in \cite{Bachas:1999um} by requiring the consistency of the effective action with the $O(\alpha'^2)$ terms of the corresponding disk-level scattering amplitude \cite{Garousi:1996ad,Hashimoto:1996kf}.  The couplings of non-constant dilaton and B-field at the order  $O(\alpha'^2)$  has been found in \cite{Garousi:2009dj} by requiring the consistency of the curvature couplings with the standard rules of linear T-duality transformations, and by the scattering amplitude. 

The CS part  which describes the coupling of the D-branes to the RR potential at order ${\cal O}(\alpha'^0)$ is given by \cite{Polchinski:1995mt,Douglas:1995bn}
\beqa
S_{CS}&=&T_{p}\int_{M^{p+1}}e^{B}C\labell{CS2}
\eeqa
where $M^{p+1}$ represents the world volume of the D$_p$-brane, $C$ is the sum over all  RR potential forms, \ie $C=\sum_{n=0}^8C^{(n)}$,  and the multiplication rule is the wedge product. The four form  is self-dual, \ie $dC^{(4)}=*(dC^{(4)})$,  and the electric components of the redundant fields $C^{(5)},\cdots, C^{(8)}$ are related to the magnetic components of the RR fields $C^{(0)},\cdots, C^{(3)}$ as $dC^{(8-n)}=*(dC^{(n)})$ for $n=0,1,2,3$. The abelian gauge field can be added to the action as $B\rightarrow B+2\pi\alpha'f$. The curvature correction to this action has been found in \cite{Green:1996dd,Cheung:1997az,Minasian:1997mm}
 by requiring that the chiral anomaly on the world volume of intersecting D-branes (I-brane) cancels with the anomalous variation of the CS action. The curvature coupling has been extended in \cite{Becker:2010ij,Garousi:2010rn} to include the   B-field at the order  $O(\alpha'^2)$   by requiring the consistency of the curvature coupling at the order  $O(\alpha'^2)$ with the  linear T-duality transformations.  

It has been shown in \cite{Garousi:2010ki} that there are another class of couplings at the order  $O(\alpha'^2)$   which involve the linear  RR field strengths $F^{(p)},\, F^{(p+2)}$ and $F^{(p+4)}$. These couplings have been found  by studying the S-matrix element of one RR and one NSNS vertex operators  \cite{Garousi:1996ad}. The couplings for $F^{(p)}$  in the string frame   are \cite{Garousi:2010ki}:
\beqa
S_{CS}&\!\!\!\!\!\supset\!\!\!\!&\frac{\pi^2\alpha'^2T_p}{2!(p-1)!}\int d^{p+1}x\,\eps^{a_0\cdots a_p}\left({ F}^{(p)}_{ia_2\cdots a_p,a}H_{a_0a_1}{}^{a,i}-\frac{1}{p}{ F}^{(p)}_{a_1a_2\cdots a_p,i}(H_{a_0a}{}^{i,a}-H_{a_0j}{}^{i,j})\right)\labell{LTdual}
\eeqa
where   commas denote partial differentiation. These couplings   are invariant under the linear T-duality and the $B$-field gauge transformations. It has been shown in \cite{Garousi:2010ki} that consistency of the above couplings with nonlinear T-duality requires one to replace the RR field strength $F^{(p)}$ with ${\cal F}^{(p)}=d{\cal C}^{p-1}$ where ${\cal C}=e^BC$. This replacement produces many couplings with nonlinear $B$-field.

The $B$-field gauge symmetry, however,  is broken by  some of the higher orders of $B$-field. The higher order  terms  which are coming from expanding ${\cal F}$, are either in terms of field strength $H$ or in terms of $B$.  The gauge symmetry is broken in the latter cases. The symmetry can be restored, however,  in the terms in which the two indices of the $B$-field are world volume, by replacing $B$ with $B+2\pi\alpha' f$.   The  terms in which   the $B$-field includes  transverse index  may be canceled in the full nonlinear T-duality invariant  action. 

A perturbative method for constructing the nonlinear T-duality invariant action  has been introduced in \cite{Garousi:2010rn}.  It works in  the following way. Consider the couplings of one RR field and $n$ number of NSNS fields at order $O(\alpha'^2)$. Suppose one makes the coupling at each order of $n$ to be invariant under "linear" T-duality transformations. By "linear" T-duality transformations we mean  that the terms with $n=1$ must be invariant under linear T-duality transformation. However, for the terms with $n>1$, one must take into account the nonlinear T-duality transformation of the terms with $m<n$ and the linear T-duality transformation of the terms with $m=n$.  Then adding  all these terms, we expect the final result to be   invariant under full nonlinear T-duality transformations  at order $O(\alpha'^2)$. As an example consider the DBI action which  is invariant under full nonlinear T-duality transformations. Upon expanding it, one finds terms with $n=1,2,\cdots$, at order $O(\alpha'^0)$. The terms with $n=1$  are  invariant under linear T-duality transformation. The terms with $n=2$ are not invariant under linear T-duality. However, when one includes the nonlinear T-duality transformations of the $n=1$ terms, one would find that the result  are invariant under T-duality. Similarly for the terms with $n>2$. 

Using the above method, one may add some new terms which are invariant under linear T-duality transformations at the level of two $B$-fields, to the action \reef{LTdual} to make it invariant under the full nonlinear T-duality transformations. In this way, one can make the resulting couplings to be invariant under the $B$-field gauge transformations \cite{Garousi:2010rn}. The gauge invariant couplings corresponding to the first term in \reef{LTdual} at the level of two $B$-fields  are then \cite{Garousi:2010ki}
\beqa
S_{CS}&\!\!\!\!\!\supset\!\!\!\!&\pi^2\alpha'^2T_{p}\int d^{p+1}x\,\eps^{a_0a_1\cdots a_{p}}\,H_{a_0a_1}{}^{a,i}\left(\frac{1}{2!(p-3)!}F^{(p-2)}_{ia_2\cdots a_{p-2}}(B_{a_{p-1}a_{p}}+2\pi\alpha'f_{a_{p-1}a_{p}})\right.\nonumber\\
&&\left.-\frac{1}{3!(p-4)!}C^{(p-3)}_{a_2\cdots a_{p-3}i}H_{a_{p-2}a_{p-1}a_{p}}+\frac{1}{2!(p-3)!}C^{(p-3)}_{a_2\cdots a_{p-2}}H_{a_{p-1}a_{p}i}\right)_{,a}\labell{first111}
\eeqa
where  using the above method, the non-gauge invariant term $B_{ia_2}F^{(p-2)}_{a_3\cdots a_{p}}$ in the expansion of ${\cal F}^{(p)}$ whose corresponding coupling is invariant under linear T-duality transformation, has been canceled  \cite{Garousi:2010rn}. 
The structure of the couplings of one RR potential $C^{(p-3)}$ and two B-fields which result from the consistency of the Chern-Simons action at order $O(\alpha'^2)$ with the linear T-duality transformations \cite{Becker:2010ij,Garousi:2010rn}, are different from the couplings in \reef{first111}. In particular the RR potential in those couplings have indices only along the brane. Restricting the RR potential $C^{(p-3)}$ to the cases that it caries transverse indices, there is no  couplings for $C^{(p-3)}$ other than those in \reef{first111}.

In this paper, we would like to confirm the couplings in  \reef{first111} by the S-matrix method. An outline of the paper is as follows: In section 2.1, using the couplings in \reef{first111}, we calculate the massless open string pole and the contact terms for the scattering amplitude of one RR and two B-fields. In section 2.2 we examine the calculation of  the S-matrix element of one RR and two NSNS vertex operators in superstring theory. In section 2.2.1, we perform the calculation in full details for $C^{(2)}_{ij}$ and expand the amplitude at low energy. We show that there is neither  contact term nor massless open string pole at order $O(\alpha'^2)$ which is consistent with the couplings \reef{first111}.  In section 2.2.2, we perform the same calculation  for $C^{(1)}_i$. We show that the massless open string pole and the contact terms of the field theory are reproduced exactly by this amplitude at order $O(\alpha'^2)$.

 \section{Scattering amplitude } \label{intro2}
A powerful method for finding the low energy field theory of the string theory is to compare the scattering amplitudes in a proposed field theory with the corresponding  amplitudes in the string theory  at the low energy. The couplings in \reef{LTdual} have been found by studying the scattering amplitude of one RR and one NSNS states. They are extended to the higher order fields in \reef{first111} by requiring the consistency of the couplings \reef{LTdual} with nonlinear T-duality. They can be confirmed by the scattering amplitude of one RR and two NSNS states. As we will see, the string theory scattering amplitude  at the low energy produces both massless open string and closed string  channels as well as some contact terms. The  closed string channels dictate the supergravity couplings in the bulk and the couplings of one RR and one NSNS states on the brane. On the other hand, the open string channels and the contact terms dictate only the couplings on the brane in which we are interested in this paper. In the next  section, we calculate the massless open string amplitude and the contact terms resulting from the couplings in \reef{first111}. 

\subsection{Field theory amplitude}

The open string channel in the scattering amplitude of one RR and two B-fields in the field theory is given by the following Feynman amplitude:
\beqa
{\cal A}_1^f&=&V_a(\veps_3,A)G_{ab}(A) V_b(A,\veps_2,\veps_1^{(n)})+(2\leftrightarrow 3)
\eeqa
\begin{center} \begin{picture}(100,100)(0,0)
\SetColor{Black}
\Line(10,40)(10,90)
\Line(10,90)(30,110)
\Line(30,110)(30,60)
\Line(30,60)(10,40)
\Text(0,108)[]{$D_4$-brane}
\Vertex(20,90){1.5}\Vertex(20,60){1.5}
\Photon(20,60)(20,90){2}{7} \Text(12,75)[]{$A$}
\Gluon(20,90)(65,90){3}{4} \Text(42.5,102.5)[]{$B_3$}
\Gluon(20,60)(65,60){3}{4}\Text(42.5,72.5)[]{$B_2$}
\Photon(20,60)(50,35){3}{4} \Text(58,45)[]{$C^{(1)}$}
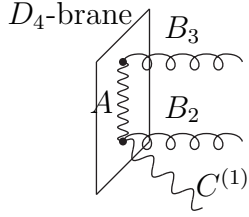
\end{picture} \\ {\sl Figure 1 :{\rm  Feynman diagram for massless open string pole.}}
\end{center}
where $A^a$ is the gauge field on the D$_p$-brane. The polarization of the RR field is given by $\veps_1^{(n)}$ and the polarizations of the B-fields are given by $\veps_2,\, \veps_3$. The on-shell conditions  are 
\beqa
p_i\inn p_i=p_i^{\mu}(\veps_i){}_{\mu \cdots}=0,\, &{\rm for}& i=1,2,3
\eeqa
To simplify the calculation we restrict the RR polarization tensor to the case that it carries only the transverse indices. Then the only non-zero vertex $ V_b(A,\veps_2,\veps_1^{(n)})$ is given by the second term in the first line of \reef{first111} for $p=4$.  The vertex is
\beqa
V_{b}(A,\veps_2,\veps_1^{(1)})&=&-2(\pi\alpha')^3T_4\eps_{a_0\cdots a_3b}(p_2\inn N\inn\veps_1)(p_2\inn V\inn H_2)^{a_0a_1}p_1^{a_2}p_2^{a_3}
\eeqa
where the  matrices  $N_{\mu\nu}$ and $V_{\mu\nu}$  project spacetime vectors into transverse and parallel subspace to the D$_p$-brane, respectively.  The gauge field propagator and the vertex $V_a(\veps_3,A)$ can be read from the DBI action \reef{DBI}, \ie
\beqa
V_a(\veps_3,A)&=&(2\pi\alpha')T_4(p_3\inn V\inn\veps_3)_a\nonumber\\
G_{ab}(A)&=&\left(\frac{-i}{T_4(2\pi\alpha')^2}\right)\frac{\eta_{ab}}{p_3\inn V\inn p_3}
\eeqa
The amplitude then becomes
\beqa
{\cal A}_1^f&=&i\left(\frac{(\pi\alpha')^2T_4}{3}\right)\frac{p_2\inn V\inn p_2\,p_2\inn N\inn\veps_1}{p_3\inn V\inn p_3}\eps_{a_0\cdots a_4}(p_3\inn V\inn\veps_3)^{a_4}p_3^{a_3}H_2^{a_0a_1a_2}+(2\leftrightarrow 3)\labell{af1}
\eeqa
This amplitude is of order $O(\alpha'^2)$ which has  six momentum in the numerator and two momentum in the denominator. 

There is a contact term at this order which is coming from the first terms in the first and second lines of \reef{first111}. They are simplify to
\beqa
{\cal A}_2^f&=&-i\frac{(\pi\alpha')^2T_4}{6}\eps_{a_0\cdots a_4}\,p_2\inn N\inn\veps_1\,p_2\inn V\inn p_2\,\veps_3^{a_3a_4}H_2^{a_0a_1a_2} +(2\leftrightarrow 3)\labell{af2}
\eeqa
Using the following identity:
\beqa
\eps_{a_0\cdots a_4}(p_3\inn V\inn H_3)^{a_3a_4}&=&\eps_{a_0\cdots a_4}\left(2(p_3\inn V\inn\veps_3)^{a_3}p_3^{a_4}+(p_3\inn V\inn p_3)\veps_3^{a_3a_4}\right)\nonumber
\eeqa
one can rewrite the sum of ${\cal A}_1^f$ and ${\cal A}_2^f$ as
\beqa
{\cal A}^f&=&-i\left(\frac{(\pi\alpha')^2T_4}{6}\right)\frac{p_2\inn V\inn p_2\,p_2\inn N\inn\veps_1}{p_3\inn V\inn p_3}\eps_{a_0\cdots a_4}(p_3\inn V\inn H_3)^{a_3a_4}H_2^{a_0a_1a_2}+(2\leftrightarrow 3)\labell{ampfield}
\eeqa
Since the result is in terms of the field strengths $H_2,\, H_3$, the amplitude satisfies the Ward identity corresponding to the B-fields. Note that neither the massless open string pole \reef{af1} nor the contact term \reef{af2} satisfy separately the Ward identity corresponding to the $\veps_3$. The massless closed string poles in which we are not interested in this paper, should also satisfy the Ward identity corresponding to the B-fields. The  amplitude \reef{ampfield}, however, does not satisfy the Ward identity corresponding to the RR field, hence, the combination of this amplitude and the massless closed string poles  should satisfy the Ward identity associated with the RR field. We will see that the string amplitude which has both open and closed string channels satisfies the Ward identities associated with the  RR and B-fields. We now turn to the string theory side and calculate the scattering amplitude of one RR and two B-field vertex operators.

 \subsection{String theory amplitude}
 
The scattering amplitude of one RR and two NSNS states has been studied in \cite{Becker:2010ij} for a particular class of terms in the amplitude to confirm some of the couplings resulting from the consistency of the Chern-Simons action at order $O(\alpha'^2)$ with the linear T-duality transformations.  In this paper, however, we are interested in the couplings in \reef{first111}, so we have to consider a different class of terms in the scattering amplitude.  In this section we study the amplitude for the general cases and then focus to the particular class of terms to examine the couplings in \reef{first111}.

In string theory, the tree level scattering amplitude of one RR and two NSNS states on the world-volume of a D$_p$-brane  is given  by the correlation function of their  corresponding vertex operators on  disk. Since the background charge of the world-sheet with topology of a disk is $Q_{\phi}=2$ one has to choose  the vertex operators in the appropriate pictures to produce the compensating charge $Q_{\phi}=-2$. One may choose the RR vertex operator in $(-1/2,-1/2)$ picture, and one of the NSNS vertex operators in $(-1,0)$ and the other one in $(0,0)$. However, in this picture the symmetry between the two NSNS is not manifest from the very beginning. After performing the correlators, one has to make more effort to rewrite the final result in a symmetric form. Alternatively, one can choose the RR vertex operator in $(-1/2,-3/2)$ picture \cite{Billo:1998vr} and the two NSNS vertex operators in $(0,0)$ picture. In this form the symmetry of the NSNS states is manifest from the beginning. We prefer to do the calculation  in the latter form. We will show that the final result, after using some identities, is independent of the choice of the picture. 

The scattering amplitude is given by the following correlation function:
\beqa
\cA&\sim&<V_{RR}^{(-1/2,-3/2)}(\veps_1^{(n)},p_1)V_{NSNS}^{(0,0)}(\veps_2,p_2)V_{NSNS}^{(0,0)}(\veps_3,p_3)>\labell{amp2}
\eeqa
Using the doubling trick \cite{Garousi:1996ad}, the vertex operators are given by the following integrals on the upper half $z$-plane:\footnote{Our conversions set $\alpha'=2$ in the string theory calculations.}
\beqa
V_{RR}^{(-1/2,-3/2)}&\!\!\!\!\!=\!\!\!\!\!&(P_-H_{1(n)}M_p)^{AB}\int d^2z_1:e^{-\phi(z_1)/2}S_A(z_1)e^{ip_1\cdot X}:e^{-3\phi(\bz_1)/2}S_B(\bz_1)e^{ip_1\cdot D\cdot  X}:\nonumber\\
V_{NSNS}^{(0,0)}&\!\!\!\!\!=\!\!\!\!\!&(\veps_2\inn D)_{\mu_3\mu_4}\int d^2z_2:(\prt X^{\mu_3}+ip_2\inn\psi\psi^{\mu_3})e^{ip_2\cdot X}:(\prt X^{\mu_4}+ip_2\inn D\inn\psi\psi^{\mu_4})e^{ip_2\cdot D\cdot X}:\nonumber\\
V_{NSNS}^{(0,0)}&\!\!\!\!\!=\!\!\!\!\!&(\veps_3\inn D)_{\mu_5\mu_6}\int d^2z_3:(\prt X^{\mu_5}+ip_3\inn\psi\psi^{\mu_5})e^{ip_3\cdot X}:(\prt X^{\mu_6}+ip_3\inn\nonumber D\inn\psi\psi^{\mu_6})e^{ip_3\cdot D\cdot X}:
\eeqa
where  the indices $A,B,\cdots$ are the Dirac spinor indices and  $P_-=\frac{1}{2}(1-\gamma_{11})$ is the chiral projection operator which makes the calculation of the gamma matrices to be with the full $32\times 32$ Dirac matrices of the ten dimensions. 
 The matrix $D^{\mu}_{\nu}$ is diagonal with $+1$ in the world volume directions and $-1$ in the transverse directions, and
\beqa
H_{1(n)}&=&\frac{1}{n!}\veps_{1\mu_1\cdots\mu_{n}}\gamma^{\mu_1}\cdots\gamma^{\mu_{n}}\nonumber\\
M_p&=&\frac{\pm 1}{(p+1)!} \eps_{a_0 \cdots a_p} \ga^{a_0} \cdots \ga^{a_p}
\eeqa
where $\eps$ is the volume $(p+1)$-form of the $D_p$-brane. It is useful to write the matrix $D_{\mu\nu}$ and the flat metric $\eta_{\mu\nu}$ in terms of the two projection operators $N_{\mu\nu}$ and $V_{\mu\nu}$, \ie
\beqa
\eta_{\mu\nu}&=&V_{\mu\nu}+N_{\mu\nu}\nonumber\\
D_{\mu\nu}&=&V_{\mu\nu}-N_{\mu\nu}
\eeqa
The components of vectors projected into each of these subspaces $N$ and $V$ or $\eta$ and $D$ are independent objects. If 1 in the chiral projection $P_-$ produces couplings for $C^{(n)}$, then the $\gamma_{11}$ produces the couplings for $C^{(10-n)}$. Hence, we consider 1 in the chiral projection and extend the result to all RR potentials.

Choosing the above integral form of the vertex operators, one has to also divide the amplitude \reef{amp2} by the volume of $SL(2,R)$ group which is the conformal symmetry of the  upper half $z$-plane. We will remove this factor after  preforming the correlators. Moreover, the overall factor of the amplitude \reef{amp2} may  be fixed by comparing the final result with field theory.

Using the standard world-sheet propagators
\beqa
<X^{\mu}(x)X^{\nu}(y)>&=&-\eta^{\mu\nu}\log(x-y)\nonumber\\
<\psi^{\mu}(x)\psi^{\nu}(y)>&=&-\frac{\eta^{\mu\nu}}{x-y}\nonumber\\
<\phi(x)\phi(y)>&=&-\log(x-y)\labell{wpro}
\eeqa
one can calculate the correlators in \reef{amp2}. The amplitude \reef{amp2}  can be written  as 
\beqa
\cA&\sim&\frac{1}{2}(H_{1(n)}M_p)^{AB}(\veps_2\inn D)_{\mu_3\mu_4}(\veps_3\inn D)_{\mu_5\mu_6}\int d^2z_1d^2z_2d^2z_3\, (z_1-\bz_1)^{-3/4}\nonumber\\
&&\times(b_1+b_2+\cdots +b_{10})^{\mu_3\mu_4\mu_5\mu_6}_{AB}\delta^{p+1}(p_1^a+p_2^a+p_3^a)+(2\leftrightarrow 3)\labell{A}
\eeqa
where the delta function which gives the conservation of the momenta along the brane is extracted  from the $X$ correlators. We have also performed the ghost correlator. The other correlators are
\beqa
(b_1)^{\mu_3\mu_4\mu_5\mu_6}_{AB}&\!\!\!\!=\!\!\!\!&<:S_A(z_1):S_B(\bz_1):>g_1^{\mu_3\mu_4\mu_5\mu_6}\nonumber\\
(b_2)^{\mu_3\mu_4\mu_5\mu_6}_{AB}&\!\!\!\!=\!\!\!\!&2(ip_2)_{\beta_1}<:S_A:S_B:\psi^{\beta_1}\psi^{\mu_3}:>g_2^{\mu_4\mu_5\mu_6}\nonumber\\
(b_3)^{\mu_3\mu_4\mu_5\mu_6}_{AB}&\!\!\!\!=\!\!\!\!&2(ip_2\inn D)_{\beta_1}<:S_A:S_B:\psi^{\beta_1}\psi^{\mu_4}:>g_3^{\mu_3\mu_5\mu_6}\nonumber\\
(b_4)^{\mu_3\mu_4\mu_5\mu_6}_{AB}&\!\!\!\!=\!\!\!\!&2(ip_2)_{\beta_1}(ip_2\inn D)_{\beta_2}<:S_A:S_B:\psi^{\beta_1}\psi^{\mu_3}:\psi^{\beta_2}\psi^{\mu_4}:>g_4^{\mu_5\mu_6}\labell{bs}\\
(b_5)^{\mu_3\mu_4\mu_5\mu_6}_{AB}&\!\!\!\!=\!\!\!\!&(ip_2)_{\beta_1}(ip_3)_{\beta_2}<:S_A:S_B:\psi^{\beta_1}\psi^{\mu_3}:\psi^{\beta_2}\psi^{\mu_5}:>g_5^{\mu_4\mu_6}\nonumber\\
(b_6)^{\mu_3\mu_4\mu_5\mu_6}_{AB}&\!\!\!\!=\!\!\!\!&2(ip_2)_{\beta_1}(ip_3\inn D)_{\beta_2}<:S_A:S_B:\psi^{\beta_1}\psi^{\mu_3}:\psi^{\beta_2}\psi^{\mu_6}:>g_6^{\mu_4\mu_5}\nonumber\\
(b_{7})^{\mu_3\mu_4\mu_5\mu_6}_{AB}&\!\!\!\!=\!\!\!\!&(ip_2\inn D)_{\beta_1}(ip_3\inn D)_{\beta_2}<:S_A:S_B:\psi^{\beta_1}\psi^{\mu_4}:\psi^{\beta_2}\psi^{\mu_6}:>g_{7}^{\mu_3\mu_5}\nonumber\\
(b_{8})^{\mu_3\mu_4\mu_5\mu_6}_{AB}&\!\!\!\!=\!\!\!\!&2(ip_2)_{\beta_1}(ip_2\inn D)_{\beta_2}(ip_3)_{\beta_3}<:S_A:S_B:\psi^{\beta_1}\psi^{\mu_3}:\psi^{\beta_2}\psi^{\mu_4}:\psi^{\beta_3}\psi^{\mu_5}:>g_{8}^{\mu_6}\nonumber\\
(b_{9})^{\mu_3\mu_4\mu_5\mu_6}_{AB}&\!\!\!\!=\!\!\!\!&2(ip_2)_{\beta_1}(ip_2\inn D)_{\beta_2}(ip_3\inn D)_{\beta_3}<:S_A:S_B:\psi^{\beta_1}\psi^{\mu_3}:\psi^{\beta_2}\psi^{\mu_4}:\psi^{\beta_3}\psi^{\mu_6}:>g_{9}^{\mu_5}\nonumber\\
(b_{10})^{\mu_3\mu_4\mu_5\mu_6}_{AB}&\!\!\!\!=\!\!\!\!&(ip_2)_{\beta_1}(ip_2\inn D)_{\beta_2}(ip_3)_{\beta_3}(ip_3\inn D)_{\beta_4}\nonumber\\
&&\qquad\qquad\qquad\times<:S_A:S_B:\psi^{\beta_1}\psi^{\mu_3}:\psi^{\beta_2}\psi^{\mu_4}:\psi^{\beta_3}\psi^{\mu_5}:\psi^{\beta_4}\psi^{\mu_6}:>g_{10}\nonumber
\eeqa
where $g$'s are  the correlators  of $X$'s. Using the propagators \reef{wpro}, one can easily calculate $g$'s. 
To find the correlator of $\psi$, we use the following Wick-like rule for the correlation function involving an arbitrary number of $\psi$'s and two $S$'s:
\beqa
 &&<:S_{A}(z_1):S_{B}(\bz_1):\psi^{\mu_1}
(z_2)\cdots
\psi^{\mu_n}(z_n):>=\labell{wicklike}\\
&&\frac{1}{2^{n/2}}
\frac{(z_1-\bz_1)^{n/2-5/4}}
{\sqrt{(z_2-z_1)(z_2-\bz_1)}\cdots\sqrt{(z_n-z_1)(z_n-\bz_1)}}\left\{(\gamma^{\mu_n\cdots\mu_1}
C^{-1})_{AB}+\right.\nonumber\\
&&\left.+\cP(z_3,z_2)\eta^{\mu_2\mu_1}(\gamma^{\mu_n\cdots\mu_3}
C^{-1})_{AB}
+\cP(z_3,z_2)\cP(z_5,z_4)\eta^{\mu_2\mu_1}\eta^{\mu_4\mu_3}
(\gamma^{\mu_n\cdots\mu_5}
C^{-1})_{AB}\right.\nonumber\\&&\left.
\qquad\qquad\qquad\qquad\qquad\qquad\qquad\qquad\qquad\qquad\qquad\qquad+\cdots\pm {\rm perms}\right\}\nonumber\eeqa where  dots mean  sum  over all possible contractions. In above equation, $\gamma^{\mu_{n}...\mu_{1}}$ is the totally antisymmetric combination of the gamma matrices and  $\cP(z_i,z_j)$ is given by the Wick-like contraction
\beqa
\cP(z_i,z_j)\eta^{\mu\nu}&=&\widehat{[\psi^{\mu}(z_i),\psi^{\nu}(z_j)]}=\eta^{\mu\nu}{\frac {(z_{i}-z_1)(z_{j}-\bz_1)+(z_{j}-z_1)(z_{i}-\bz_1)}{(z_{i}-z_{j})(z_1-\bz_1)}}\labell{ppp}
\eeqa
The Wick-like rule has been found in \cite{Liu:2001qa} for the case that  $\psi$'s are   set at  different points in the real axis. It has been extended in \cite{Garousi:2008ge} to the case that $\psi$'s in the real axis appear as currents, \ie two $\psi$'s appear in one point. In that case, the only subtlety in using the  Wick-like rule    is that one must not consider the Wick-like contraction for the two $\psi$'s in one  current. In above we have extended further the formula to the case that the $\psi$'s are in the complex plane. See \cite{Hartl:2009yf, Haertl:2010ks}, for the correlation function of $\psi$'s with four and more spin operators.   

 Combining the gamma matrices coming from the  correlation \reef{wicklike} with the gamma matrices in \reef{A}, one finds  the following  trace:
 \beqa
T(n,p,m)& =&(H_{1(n)}M_p)^{AB}(\gamma^{\alpha_1\cdots \alpha_m}C^{-1})_{AB}A_{[\alpha_1\cdots \alpha_m]}\labell{relation1}\\
& =&\frac{1}{n!(p+1)!}\veps_{1\nu_1\cdots \nu_{n}}\eps_{a_0\cdots a_p}A_{[\alpha_1\cdots \alpha_m]}\Tr(\gamma^{\nu_1}\cdots \gamma^{\nu_{n}}\gamma^{a_0}\cdots\gamma^{a_p}\gamma^{\alpha_1\cdots \alpha_m})\nonumber
 \eeqa
where $A_{[\alpha_1\cdots \alpha_m]}$ is an antisymmetric combination of the momenta and/or the polarizations of the NSNS states. For example the correlator  $b_{10}$ in \reef{bs} produces terms with $m=0,2,4,6,8$. For $m=8$ it is
\beqa
A_{[\alpha_1\cdots \alpha_8]}&=&(\veps_3\inn D)_{[\alpha_3\alpha_1}(\veps_2\inn D)_{\alpha_7\alpha_5}(ip_2)_{\alpha_8}(ip_2\inn D)_{\alpha_6}(ip_3)_{\alpha_4}(ip_3\inn D)_{\alpha_2]}\labell{Am8}
\eeqa
The trace \reef{relation1} can be evaluated for specific values of $n$. One can  verify that the amplitude is non-zero only  for $n=p-3,\, n=p-1,\, n=p+1,\, n=p+3,\, n=p+5$. 

To examine the couplings in \reef{first111} with the S-matrix element \reef{amp2}, we consider in this paper only the case 
\beqa
n&=&p-3\labell{np3}
\eeqa
In this case, the trace relation \reef{relation1} gives non-zero result only for $m\geq 4$.  One immediately concludes that  $b_1,\, b_2$ and $b_3$ in \reef{bs} have no contribution to the amplitude.
There are still a lot of  terms with  non-zero contributions.  For ease of calculation we restrict the RR polarization tensor to some specific cases as in the field theory calculation. We consider the cases that the  non-vanishing components of the RR polarization to be $\veps_1^{ij}$ and $\veps_1^i$.  Let us begin with the RR polarization  $\veps_1^{ij}$.

\subsubsection{$\veps_1^{ij}$ case} \label{5,8}

In this case $n=2$, and from the relation \reef{np3} one gets $p=5$. Since the indices of the RR potential are transverse  and the indices of the volume form are the world-volume indices, the trace  \reef{relation1} is non-zero only for $m=8$. The trace for $m=8$ becomes 
\beqa
T(2,5,8)&=&32\frac{8!}{2!6!}\veps_1^{ij}\eps^{a_0\cdots a_5}A_{[ija_0\cdots a_5]}
\eeqa
where 32 is the trace of the $32\times 32$ identity matrix.

Since $m=8$,  only $b_{10}$ has non-zero contribution to the amplitude \reef{amp2}. The $X$ correlator in  $b_{10}$ is 
\beqa
g_{10}&\!\!\!\!=\!\!\!\!&|z_{12}|^{2p_1\cdot p_2}|z_{13}|^{2p_1\cdot p_3}|z_{23}|^{2p_2\cdot p_3}|z_{1\bar2}|^{2p_1\cdot D\cdot p_2}|z_{1\bar3}|^{2p_1\cdot D\cdot p_3}|z_{2\bar3}|^{2p_2\cdot D\cdot p_3} \nonumber\\
&&\times(z_{1\bar1})^{p_1\cdot D\cdot p_1}(z_{2\bar2})^{p_2\cdot D\cdot p_2}(z_{3\bar3})^{p_3\cdot D\cdot p_3}(i)^{p_1\cdot D\cdot p_1+p_2\cdot D\cdot p_2+p_3\cdot D\cdot p_3}\equiv K\labell{K}
\eeqa
where $z_{ij}=z_i-z_j$ and $z_{i\bar j}=z_i-\bz_j$. We have added the phase factor to make it real. The above function  appears in all other $X$ correlators in \reef{bs}. 

Replacing in \reef{A} the above  $X$-correlator and the $\psi$-correlator  from \reef{wicklike}, one finds
 \beqa
{\cal A}&\sim&\frac{1}{2}\frac{8!}{6!} \veps_1^{ij}\eps^{a_0\cdots a_5}A_{[ija_0\cdots a_5]}\cI \labell{amp3}
\eeqa
where $A_{[ija_0\cdots a_5]}$ is given in \reef{Am8}. 
The conservation of momentum is understood   in the above and in all other amplitudes in this paper. The integral in the above amplitude is 
\beqa
\cI&=&\int d^2z_1d^2z_2d^2z_3\,\frac{ \left(z_{1\bar1}\right)^2 K}{|z_{12}|^2 |z_{13}|^2 |z_{2\bar1}|^2 |z_{3\bar1}|^2 }\labell{I}
\eeqa
which is real as expected. To simplify the kinematic factor in \reef{amp3}, we note that  the indices of the RR polarization and the volume form are each antisymmetric. Hence, there are 28 different terms, however, using the definition of matrix D
\beqa
D^{\mu i}=-\delta^{\mu i}, &{\rm and}& D^{\mu a}=\delta ^{\mu a}
\eeqa
24 of them are zero, and the rest are identical. The  amplitude \reef{amp3} then becomes
\beqa
{\cal A}&\sim&4 (\veps_1)_{ij}\eps_{a_0\cdots a_5}p_2^{a_0}p_3^{a_1}p_2^ip_3^j\veps_2^{a_2a_3}\veps_3^{a_4a_5}\cI+(2\leftrightarrow 3)\labell{amp5}
\eeqa
 It can   be seen that the amplitude satisfies the Ward identities corresponding to the $B$-fields. Using the fact that the only non-zero components of the RR polarization is $(\veps_1)_{ij}$ one  observes that the amplitude also satisfies the Ward identity corresponding to the RR gauge field, \ie the amplitude can be written in terms of the RR field strength $F_{a_0ij}$.

We now calculate the integral \reef{I}. Using the conservation of the momentum, \ie
\beqa
(p_1+p_2+p_3)\inn V_{\mu}&=&0\labell{cons}
\eeqa
one can easily check that the integrand is  invariant under the $SL(2,R)$ transformation. So we can map the results to disk with unit radius. The map from upper half plane to unit disk is
\beqa
z\rightarrow -i\frac{z-1}{z+1}
\eeqa
Under this map, one has
\beqa
z_{i\bar j}&\rightarrow &-\frac{-2i}{(z_i+1)(\bz_j+1)}(1-z_i\bz_j)\nonumber\\
z_{ij}&\rightarrow &\frac{-2i}{(z_i+1)(z_j+1)}z_{ij}\nonumber\\
d^2z_i&\rightarrow &\frac{4}{(z_i+1)(\bz_i+1)}d^2z_i
\eeqa
The factors like  $\frac{-2i}{(z_i+1)(\bz_j+1)}$ are canceled because the integrand is $SL(2,R)$ invariant. So to map the integral \reef{I} to unit disk one has to use the following replacement:
\beqa
z_{i\bar j}&\rightarrow &-(1-z_i\bz_j)\nonumber\\
z_{ij}&\rightarrow &z_{ij}\nonumber\\
z_{\bar i\bar j}&\rightarrow &-z_{\bar i\bar j}\nonumber\\
z_{\bar i  j}&\rightarrow &(1-\bz_iz_j)\nonumber
\eeqa
Obviously the result is still  $SL(2,R)$ invariant. To fix this symmetry, we then set \cite{Craps:1998fn}
\beqa
z_1=0,&{\rm and}& z_2=\bz_2=r_2\labell{fix}
\eeqa
Under this fixing the measure changes as
\beqa
d^2z_1d^2z_2d^2z_3\rightarrow r_2dr_2\,r_3dr_3\,d\theta,&&0<r_2,r_3<1,\,0<\theta<2\pi
\eeqa
where we have chosen  the polar coordinate $z_3=r_3e^{i\theta}$. The integral \reef{I} then becomes
\beqa
\cI&=&\int_0^1dr_2\int_0^1dr_3\int_0^{2\pi}d\theta\,\frac{ \tK}{r_2r_3}\labell{amp4}
\eeqa
where $\tK$ is 
\beqa
\tK&\!\!\!\!=\!\!\!\!&\ {r_2}^{2p_1\cdot p_2}\ {r_3}^{2 p_1\cdot p_3} (1-{r_2}^2)^{p_2\cdot D\cdot p_2}(1-{r_3}^2)^{p_3\cdot D\cdot p_3}\nonumber\\
&&\left.\times |r_2-r_3 e^{i \t}|^{2 p_2\cdot p_3}|1-r_2 r_3 e^{i \t}|^{2 p_2\cdot D\cdot p_3}\right. \labell{KK}
\eeqa

By looking at $\tK$, one realizes that there are 6 independent Mandelstam variables in the general case of three closed string amplitude, \ie
\beqa
p_1\inn p_2,\,\, p_1\inn p_3,\,\,p_2\inn p_3,\,\, p_2\inn D\inn p_2,\,\,p_3\inn D\inn p_3,\,\,p_2\inn D\inn p_3\labell{Mandel}
\eeqa
This number is not the same as the physical open or closed string channels. There are 4 closed string channels $p_1\inn p_2$, $p_1\inn p_3$, $p_2\inn p_3$ and $(p_1+p_2+p_3)^2$, and there are 3 open string channels $p_1\inn D\inn p_1$, $p_2\inn D\inn p_2$ and $p_3\inn D\inn p_3$. The physical channels $(p_1+p_2+p_3)^2$ and $p_1\inn D\inn p_1$ can be written in terms of the above independent variables as
\beqa
(p_1+p_2+p_3)^2&=&2p_1\inn p_2+2p_1\inn p_3+2p_2\inn p_3\nonumber\\
p_1\inn D\inn p_1&=&p_2\inn D\inn p_2+p_3\inn D\inn p_3+2p_2\inn p_3+2p_2\inn D\inn p_3
\eeqa
There is no open or closed string channel corresponding to the Mandelstam variable $p_2\inn\inn D\inn p_3$.

In the  disk level amplitude of  the color-ordered  $N$  open string states, there are  $N(N-3)/2$ independent Mandelstam variables and there are the  same number of open string channels. However, not all the Mandelstam variables in the  physical open string channels  are  the independent  variables that appear naturally in the amplitude. There are Mandelstam variables for which there are no open string channel. Hence, one may set them to zero to simplify the calculations. Using this observation, the scattering amplitude of $N$-point function has been studied in a restricted kinematic setup in \cite{Garousi:2010er}. This makes the calculation of the integrals involved to be much easier to perform. 

To make easier the calculation of the integral over the $\theta$-coordinate in \reef{amp4}, we here also work in a restricted kinematic setup. Examining  the Feynman diagrams involved, one can easily verify that   the amplitude considered in this paper has no massless pole in the $p_2\inn p_3$-channel. On the other hand the kinematic factor $\veps_1^{ij}\eps^{a_0\cdots a_5}A_{[ija_0\cdots a_5]}$ does not have any term proportional to   $ p_2\inn p_3$. That means the integral in \reef{amp4} has no $1/(p_2\inn p_3)$ pole. 
Moreover,  there is no closed or open string channel corresponding to the Mandelstam variable $p_2\inn D\inn p_3$, hence,   
 it is safe to restrict the Mandelstam variables in \reef{KK} to 
\beqa
 p_2\inn D\inn p_3=0, &{\rm and }& 
 p_2\inn p_3=0\labell{kin}
 \eeqa
Let us  first consider only the constraint $p_2\inn p_3=0$. This makes  the integral over  $\theta$ in \reef{amp4} to be 
\beqa
\cI&=&2\pi\int_0^1dr_2\int_0^1dr_3 {r_2}^{2p_1\cdot p_2-1}\ {r_3}^{2 p_1\cdot p_3-1} (1-{r_2}^2)^{p_2\cdot D\cdot p_2}(1-{r_3}^2)^{p_3\cdot D\cdot p_3}\nonumber\\
&&\qquad\qquad\qquad\qquad\times{}_2F_1\bigg[{-p_2\inn D\inn p_3,\ -p_2\inn D\inn p_3 \atop 1}\ ; \ r_2^2r_3^2\bigg]\labell{amp41}
\eeqa
where we have used the following relation \cite{ISG}:
\beqa
\int_0^{2\pi}d\theta\frac{\cos(n\theta)}{(1+a^2-2a\cos(\theta))^b}&=&2\pi a^n\frac{\Ga(b+n)}{n!\Ga(b)}{}_2F_1\bigg[{b, \ n+b \atop n+1}\ ;\ a^2\bigg]
\eeqa
where $|a|<1$. Then using the integral representation of the generalized hypergeometric function ${}_pF_q$ (see Appendix),
one finds that the integral over  $r_2$ and $r_3$ gives
\beqa
\cI&=&\frac{\pi}{2}B(p_1\inn p_2,1+p_2\inn D\inn p_2)B(p_1\inn p_3,1+p_3\inn D\inn p_3)\nonumber\\
&&\qquad\qquad\qquad\qquad\qquad\times{}_4F_3\bigg[{p_1\inn p_2,\ p_1\inn p_3, \ -p_2\inn D\inn p_3  \ -p_2\inn D\inn p_3  \atop 1+p_1\inn p_2+p_2\inn D\inn p_2, \ 1+p_1\inn p_3+p_3\inn D\inn p_3, \ 1}\ ;\ 1\bigg]\nonumber
\eeqa
As we have anticipated above, there is no closed or open string channels in $p_2\inn D\inn p_3$. So it is consistent to use the constraint \reef{kin} under which the above integral simplifies to
\beqa
\cI&=&\frac{\pi}{2}B(p_1\inn p_2,1+p_2\inn D\inn p_2)B(p_1\inn p_3,1+p_3\inn D\inn p_3)\labell{amp51}
\eeqa
To study the low energy limit of the amplitude \reef{amp5}, we expand $\cI$  at the low energy, \ie 
\beqa
\cI&=&\frac{\pi}{2}\left(\frac{1}{p_1\inn p_2 p_1\inn p_3}-\frac{\pi^2}{6}\bigg[\frac{p_2\inn D\inn p_2}{p_1\inn p_3}+\frac{p_3\inn D\inn p_3}{p_1\inn p_2}\bigg]+\cdots \right)\labell{amp6}
\eeqa
The first term is a double massless closed string pole and the second terms are simple massless closed string poles.  The dots in the above expansion include terms with four and higher order of momenta. We have seen in the field theory calculation in the previous section that there is no massless open string channel for $\veps_1^{ij}$. This  is consistent with the expansion of \reef{amp6} which has no massless open string channel.

Before leaving this section,  we note that the double pole may seems to be unphysical pole, \ie there is no Feynman diagram corresponding to this pole. However, one can rearrange it in terms of the physical massless closed string double pole  as 
\beqa
\frac{1}{p_1\inn p_2 p_1\inn p_3}=\frac{1}{p_1\inn p_2(p_1\inn p_2+p_1\inn p_3)}+\frac{1}{p_1\inn p_3(p_1\inn p_2+p_1\inn p_3)}\labell{decom}
\eeqa
and the pole $1/(p_1\inn p_2+p_1\inn p_3)$ is in fact $ 2/(p_1+ p_2+ p_3)^2$ in which the constraint \reef{kin} has been used. 
Similar decomposition can be done for massive poles of the amplitude \reef{amp5}. So the amplitude \reef{amp5} has three closed string channels $p_1\inn p_2$, $p_1\inn p_3$, $(p_1\inn p_2+p_1\inn p_3)$ and three open string channels $p_2\inn D\inn p_2$, $p_3\inn D\inn p_3$,  $p_1\inn D\inn p_1$, however, there are no massless  poles in the latter channels.

\subsubsection{$\veps_1^i$ case} \label{4,6}

In this case $n=1$, and from the relation \reef{np3} one finds $p=4$. Since the index of the RR potential is transverse and the indices of the volume form are the world-volume indices, one finds that the trace relation \reef{relation1} is non-zero only for $m=6$. The trace in this case becomes
\beqa
T(1,4,6)&=&32\frac{6!}{5!}\veps_1^i\eps^{a_0\cdots a_4}A_{[ia_0\cdots a_4]}
\eeqa

Since $m=6$, the $\psi$ correlators in $b_{8},\, b_{9}$ and $b_{10}$ have non-zero contributions. For $b_{8},\, b_{9}$ the first term in \reef{wicklike} and for $b_{10}$ the second terms in \reef{wicklike} have non-zero contributions. The $X$ correlators in $b_{8},\, b_{9},\, b_{10}$ are
\beqa
g_8^{\mu_6}&=& i\left(\frac{p_1^{\mu _6}}{z_{1\bar3}}+\frac{p_2^{\mu _6}}{z_{2\bar3}}+\frac{p_3^{\mu _6}}{z_{3\bar3}}+\frac{(p_1\inn D)^{\mu _6}}{{z}_{\bar1\bar3}}+\frac{(p_2\inn D)^{\mu _6}}{{z}_{\bar2\bar3}}\right)K\nonumber\\
g_9^{\mu_5}&=&  i\left(\frac{p_1^{\mu _5}}{z_{13}}+\frac{p_2^{\mu _5}}{z_{23}}+\frac{(p_3\inn D)^{\mu _5}}{z_{\bar3 3}}+\frac{(p_1\inn D)^{\mu _5}}{{z}_{\bar1 3}}+\frac{(p_2\inn D)^{\mu _5}}{{z}_{\bar2 3}}\right)K\nonumber\\
g_{10}&=&K
\eeqa
Where $K$ is given in \reef{K}. Using the on-shell condition $\veps_2\cdot p_2=\veps_3\cdot p_3=0$ and conservation of momentum along the world volume, \ie
\beqa
p_1+p_1\inn D+p_2+p_2\inn D+p_3+p_3\inn D&=&0
\eeqa
one can rewrite $g_{8},\, g_{9}$ as 
\beqa
g_8^{\mu_6}&=&  \frac{iK}{z_{3\bar3}}\left(\frac{p_1^{\mu _6}z_{31}}{z_{1\bar3}}+\frac{p_2^{\mu _6}z_{32}}{z_{2\bar3}}+\frac{(p_1\inn D)^{\mu _6}z_{3\bar1}}{{z}_{\bar1\bar3}}+\frac{(p_2\inn D)^{\mu _6}z_{3\bar2}}{{z}_{\bar2\bar3}}\right)\nonumber\\
g_9^{\mu_5}&=&  \frac{iK}{z_{\bar3 3}}\left(\frac{p_1^{\mu _5}z_{\bar3 1}}{z_{13}}+\frac{p_2^{\mu _5}z_{\bar3 2}}{z_{23}}+\frac{(p_1\inn D)^{\mu _5}z_{\bar3\bar1}}{{z}_{\bar1 3}}+\frac{(p_2\inn D)^{\mu _5}z_{\bar3\bar2}}{{z}_{\bar2 3}}\right)\nonumber
\eeqa
 The reason for this arrangement is that all terms behave similarly  under the $SL(2,R)$ transformation. Writing the sub-amplitudes  ${\cal A}_i$ in \reef{A} corresponding to $b_i$,  the sub-amplitudes corresponding to $b_{8},\, b_9$ are
\beqa
{\cal A}_8&\!\!\!\!\!\sim\!\!\!\!\!& 4\frac{6!}{5!}\veps_1^i\eps^{a_0\cdots a_4}(D\inn\veps_3^T)_{\mu_6[i}(\veps_2\inn D)_{a_3a_1}(p_2)_{a_4}(p_2\inn D)_{a_2}(p_3)_{a_0]}\nonumber\\
&&\int d^2z_1d^2z_2d^2z_3\frac{z_{1\bar1}K}{|z_{21}|^2|z_{2\bar1}|^2z_{31}z_{3\bar1}z_{3\bar3}}\left(\frac{p_1^{\mu _6}z_{31}}{z_{1\bar3}}+\frac{p_2^{\mu _6}z_{32}}{z_{2\bar3}}+\frac{(p_1\inn D)^{\mu _6}z_{3\bar1}}{{z}_{\bar1\bar3}}+\frac{(p_2\inn D)^{\mu _6}z_{3\bar2}}{{z}_{\bar2\bar3}}\right)\nonumber\\
{\cal A}_9&\!\!\!\!\!\sim\!\!\!\!\!& 4\frac{6!}{5!}\veps_1^i\eps^{a_0\cdots a_4}(\veps_3\inn D)_{\mu_5[i}(\veps_2\inn D)_{a_3a_1}(p_2)_{a_4}(p_2\inn D)_{a_2}(p_3\inn D)_{a_0]}\nonumber\\
&&\int d^2z_1d^2z_2d^2z_3\frac{z_{1\bar1}K}{|z_{21}|^2|z_{2\bar1}|^2z_{\bar31}z_{\bar3\bar1}z_{\bar3 3}}\left(\frac{p_1^{\mu _5}z_{\bar3 1}}{z_{13}}+\frac{p_2^{\mu _5}z_{\bar3 2}}{z_{23}}+\frac{(p_1\inn D)^{\mu _5}z_{\bar3\bar1}}{{z}_{\bar1 3}}+\frac{(p_2\inn D)^{\mu _5}z_{\bar3\bar2}}{{z}_{\bar2 3}}\right)\nonumber
\eeqa
 To calculate the kinematic factors, we note that there are 6 different terms, however, 4 of them are zero and the other two terms are equal. Using $D_{\mu a}=\delta_{\mu a}$, the amplitudes then become
\beqa
{\cal A}_8&\!\!\!\!\!\sim\!\!\!\!\!& -8\veps_1^i\eps^{a_0\cdots a_4}(D\inn\veps_3^T)_{\mu_6a_3}(\veps_2)_{a_1a_4}(p_2)_{i}(p_2)_{a_2}(p_3)_{a_0}\labell{A89}\\
&&\int d^2z_1d^2z_2d^2z_3\frac{z_{1\bar1}K}{|z_{21}|^2|z_{2\bar1}|^2z_{31}z_{3\bar1}z_{3\bar3}}\left(\frac{p_1^{\mu _6}z_{31}}{z_{1\bar3}}+\frac{p_2^{\mu _6}z_{32}}{z_{2\bar3}}+\frac{(p_1\inn D)^{\mu _6}z_{3\bar1}}{{z}_{\bar1\bar3}}+\frac{(p_2\inn D)^{\mu _6}z_{3\bar2}}{{z}_{\bar2\bar3}}\right)\nonumber\\
{\cal A}_9&\!\!\!\!\!\sim\!\!\!\!\!& -8\veps_1^i\eps^{a_0\cdots a_4}(\veps_3)_{\mu_5a_3}(\veps_2)_{a_1a_4}(p_2)_{i}(p_2)_{a_2}(p_3)_{a_0}\nonumber\\
&&\int d^2z_1d^2z_2d^2z_3\frac{z_{1\bar1}K}{|z_{21}|^2|z_{2\bar1}|^2z_{\bar31}z_{\bar3\bar1}z_{\bar3 3}}\left(\frac{p_1^{\mu _5}z_{\bar3 1}}{z_{13}}+\frac{p_2^{\mu _5}z_{\bar3 2}}{z_{23}}+\frac{(p_1\inn D)^{\mu _5}z_{\bar3\bar1}}{{z}_{\bar1 3}}+\frac{(p_2\inn D)^{\mu _5}z_{\bar3\bar2}}{{z}_{\bar2 3}}\right)\nonumber
\eeqa

There are similar sub-amplitudes as above  in  the $(2\leftrightarrow 3)$ part of the amplitude \reef{A}. The above amplitudes are zero for symmetric polarization tensor $\veps_2$. It produces terms which contain $(p\veps_3)_{\mu}$.  In the  $p_1\inn V\inn \veps_3$ part of the above amplitudes, one can use the conservation of momentum along the brane \reef{cons}  to write them in terms of $p_2\inn V\inn \veps_3$ and $p_2\inn V\inn \veps_3$.
 We will see shortly that there are other $p_2\veps_3$ and $p_3\veps_3$ contributions from  $b_{10}$. However,   there is no contribution to the $p_1\inn N\inn\veps_3$, $p_2^i\,p_2\inn N\inn\veps_3$ and $p_2^i\,p_2\inn V\inn\veps_3$ from $b_{10}$. So let us consider these structures in the above amplitudes.

The $p_1\inn N\inn\veps_3$ structure of the amplitude  is zero for symmetric polarization tensor $\veps_3$, and for antisymmetric  tensor $\veps_3$ it is 
\beqa
{\cal A}(p_1\inn N\inn\veps_3)&\sim&-16\veps_1^i\eps^{a_0\cdots a_4}(p_1\inn N\inn\veps_3)_{a_3}(\veps_2)_{a_1a_4}(p_2)_{i}(p_2)_{a_2}(p_3)_{a_0}\cI_1\labell{I1}
\eeqa
where
\beqa
\cI_1
&=&\int d^2z_1d^2z_2d^2z_3\frac{z_{1\bar1}K}{|z_{21}|^2|z_{2\bar1}|^2z_{3\bar3}}\left(\frac{1}{z_{1\bar3}z_{3\bar1}}-\frac{1}{z_{\bar1\bar3}z_{31}}\right)\nonumber
\eeqa
The integrand is real as expected. One can easily verify that the integral is the one appears in the previous section, \ie
\beqa
\cI_1&=&-\cI \labell{iden0}
\eeqa
The $p_2^i\,p_2\inn N\inn\veps_3$ structure is 
\beqa
{\cal A}(p_2\inn N\inn\veps_3,p_2\inn N\inn\veps_1)&\sim &8\veps_1^i\eps^{a_0\cdots a_4}(p_2\inn N\inn\veps_3)_{a_3}(\veps_2)_{a_1a_4}(p_2)_{i}(p_2)_{a_2}(p_3)_{a_0}\cI_2\labell{I2}
\eeqa
where
\beqa
\cI_2
&=&\int d^2z_1d^2z_2d^2z_3\frac{z_{1\bar1}(|z_{32}|^2-|z_{\bar3 2}|^2)K}{|z_{21}|^2|z_{2\bar1}|^2z_{3\bar3}}\left(\pm\frac{1}{z_{31}z_{3\bar 1}z_{2\bar3}z_{\bar2\bar3}}+\frac{1}{z_{\bar3\bar1}z_{\bar3 1}z_{23}z_{\bar2 3}}\right)\nonumber
\eeqa
where the plus sign is for antisymmetric  tensor $\veps_3$ and the minus sign is for the symmetric tensor. The integrand is pure imaginary for symmetric  and is real for antisymmetric case. 
The  $p_2^i\,p_2\inn V\inn\veps_3$ structure  is 
\beqa
{\cal A}(p_2\inn V\inn\veps_3,p_2\inn N\inn\veps_1)&\sim &-8\veps_1^i\eps^{a_0\cdots a_4}(p_2\inn V\inn\veps_3)_{a_3}(\veps_2)_{a_1a_4}(p_2)_{i}(p_2)_{a_2}(p_3)_{a_0}\cI_3\labell{I3}
\eeqa
where
\beqa
\cI_3
&=&\int d^2z_1d^2z_2d^2z_3\frac{z_{1\bar1}K}{|z_{21}|^2|z_{2\bar1}|^2z_{3\bar3}}\left(\frac{1\pm 1}{z_{1\bar3}z_{3\bar1}}+\frac{1\pm 1}{z_{\bar1\bar3}z_{31}}\pm\frac{|z_{32}|^2+|z_{\bar3 2}|^2}{z_{31}z_{3\bar 1}z_{2\bar3}z_{\bar2\bar3}}+\frac{|z_{32}|^2+|z_{\bar3 2}|^2}{z_{\bar3\bar1}z_{\bar3 1}z_{23}z_{\bar2 3}}\right)\nonumber
\eeqa
where the plus sign is for antisymmetric   and minus sign is for the symmetric tensor $\veps_3$. The integrand is pure imaginary for symmetric  and is real for antisymmetric case. The RR tensor in above amplitudes can be written in terms of $F_{a_0i}$, so they are invariant under the RR gauge transformation. The $\veps_2$ can also be written in terms of the field strength. So they satisfy the Ward identity associated with $\veps_2$. 
However, they do not satisfy the Ward identity associated with $\veps_3$. We will see later that the combination of these terms and some other terms in $b_{10}$ satisfy the Ward identity.

Now let us calculate the  ${\cal A}_{10}$ part of the amplitude. The $\psi$ correlators \reef{wicklike} for  $b_{10}$ gives 24 different contractions. There are 6 contractions in which two indices of the NSNS polarization tensors contract with each other, \eg $(\veps_2\inn D\inn\veps_3)_{\mu_3\mu_5}$. They  give zero result because three indices of the factor $(ip_2)_{\beta_1}(ip_2\inn D)_{\beta_2}(ip_3)_{\beta_3}(ip_3\inn D)_{\beta_4}$ must then contract with the world-volume tensor which is zero. There are  6 contractions where two momenta contract with each other, \eg $p_2\inn D\inn p_2$. There are 6 contractions which produce structure $p\veps_3$. These terms  have similar  structure as terms in \reef{A89}. There are  another 6 contractions which produce  structure $p\veps_2$. They have similar structure as the $(2\leftrightarrow 3)$ partner of \reef{A89}. The contractions which produce structure $p\veps_3$  give the following contribution to the amplitude:
\beqa
{\cal A}_{10}(p\veps_3)&\sim&2\frac{6!}{5!}\veps_1^i\eps^{a_0\cdots a_4}\int d^2z_1d^2z_2d^2z_3\frac{z_{1\bar1}^2K}{|z_{21}|^2|z_{31}|^2|z_{2\bar 1}|^2|z_{3\bar 1}|^2}\nonumber\\
&& \bigg(\cP(z_2,z_3) A_{1[ia_0\cdots a_4]} +\cP(\bar{z}_2,z_3)A_{2[ia_0\cdots a_4]}  
+\cP(z_3,\bar{z}_3)  A_{3[ia_0\cdots a_4]}\nonumber\\ 
&&-\cP(z_2,\bar{z}_3) A_{4[ia_0\cdots a_4]} 
+\cP(z_3,{\bz}_3)A_{5[ia_0\cdots a_4]}  -\cP(\bar{z}_2,\bar{z}_3)A_{6[ia_0\cdots a_4]}  \bigg) \labell{amp12}
\eeqa
where $\cP(z_i,z_j)$ is given in \reef{ppp} and
\beqa
A_{1[ia_0\cdots a_4]}&=&(p_2\inn\veps_3\inn D)_{[a_4}(\veps_2\inn D)_{a_2a_3}(p_2\inn D)_{a_0}(p_3)_{i}(p_3\inn D)_{a_1]}\nonumber\\
A_{2[ia_0\cdots a_4]}&=&(p_2\inn D\inn\veps_3\inn D)_{[a_4}(\veps_2\inn D)_{a_2a_3}(p_2)_{i}(p_3)_{a_0}(p_3\inn D)_{a_1]}\nonumber\\
A_{3[ia_0\cdots a_4]}&=&(p_3\inn D\inn\veps_3\inn D)_{[a_4}(\veps_2\inn D)_{a_2a_3}(p_2)_i(p_2\inn D)_{a_1}(p_3)_{a_0]}\nonumber\\
A_{4[ia_0\cdots a_4]}&=&(p_2\inn D\inn\veps_3^T)_{[a_4}(\veps_2\inn D)_{a_2a_3}(p_2\inn D)_{a_0}(p_3)_{i}(p_3\inn D)_{a_1]}\nonumber\\
A_{5[ia_0\cdots a_4]}&=&(p_3\inn D\inn\veps_3^T)_{[a_4}(\veps_2\inn D)_{a_2a_3}(p_2)_i(p_2\inn D)_{a_0}(p_3\inn D)_{a_1]}\nonumber\\
A_{6[ia_0\cdots a_4]}&=&(p_2\inn\veps_3^T)_{[a_4}(\veps_2\inn D)_{a_2a_3}(p_2)_{i}(p_3)_{a_0}(p_3\inn D)_{a_1]}\nonumber
\eeqa
These factors  are zero for symmetric polarization tensor $\veps_2$.

Using the fact that above kinematic factors contract with $\veps_1^i\eps^{a_0\cdots a_4}$, one observes  that there are 6 different terms in each case, however, 4 of them are zero and the other two terms are equal. They simplify as
\beqa
A_{1[ia_0\cdots a_4]}&=&\frac{1}{3}(p_2\inn\veps_3)_{a_4}(\veps_2)_{a_2a_3}(p_2)_{a_0}(p_3)_{i}(p_3)_{a_1}\nonumber\\
A_{2[ia_0\cdots a_4]}&=&-\frac{1}{3}(p_2\inn D\inn\veps_3)_{a_4}(\veps_2)_{a_2a_3}(p_2)_{a_0}(p_3)_{i}(p_3)_{a_1}\nonumber\\
A_{3[ia_0\cdots a_4]}&=&\frac{1}{3}(p_3\inn D\inn\veps_3)_{a_4}(\veps_2)_{a_2a_3}(p_2)_i(p_2)_{a_1}(p_3)_{a_0}\nonumber\\
A_{4[ia_0\cdots a_4]}&=&\frac{1}{3}(p_2\inn D\inn\veps_3^T)_{a_4}(\veps_2)_{a_2a_3}(p_2)_{a_0}(p_3)_{i}(p_3)_{a_1}\nonumber\\
A_{5[ia_0\cdots a_4]}&=&-\frac{1}{3}(p_3\inn D\inn\veps_3^T)_{a_4}(\veps_2)_{a_2a_3}(p_2)_i(p_2)_{a_1}(p_3)_{a_0}\nonumber\\
A_{6[ia_0\cdots a_4]}&=&-\frac{1}{3}(p_2\inn\veps_3^T)_{a_4}(\veps_2)_{a_2a_3}(p_2)_{a_0}(p_3)_{i}(p_3)_{a_1}\nonumber
\eeqa

All terms in  \reef{A89} produce structure $p_2\inn N\inn\veps_1$. On the other hand, only $A_3, A_{5}$ in the above amplitude produce this structure. The other terms produce $p_3\inn N\inn\veps_1$ structure. Since there is no momentum conservation in the transverse subspace the $p_2\inn N\inn \veps_1$ and $p_3\inn N\inn \veps_1$ are independent. The $A_3, A_{5}$ in the above amplitude have contribution to the $p_3\inn V\inn\veps_3$ structure of the amplitudes \reef{A89}. Adding them, one finds that  the $p_3\inn V\inn\veps_3$ structure of the amplitude \reef{A} is zero for symmetric tensor $\veps_3$, and for antisymmetric tensor it is 
\beqa
{\cal A}(p_3\inn V\inn\veps_3,p_2\inn N\inn\veps_1)&\sim &32\veps_1^i\eps^{a_0\cdots a_4}(p_3\inn V\inn\veps_3)_{a_3}(\veps_2)_{a_1a_4}(p_2)_{i}(p_2)_{a_2}(p_3)_{a_0}\cI_4\labell{I4}
\eeqa
where
\beqa
\cI_4
&=&\int d^2z_1d^2z_2d^2z_3\frac{z_{1\bar1}^2\cP(z_3,\bz_3)K}{|z_{21}|^2|z_{31}|^2|z_{2\bar 1}|^2|z_{3\bar 1}|^2}\labell{I44}
\eeqa

The other terms in \reef{amp12} produce structure $p_3\inn N\inn\veps_1$. They are
\beqa 
{\cal A}(p_2\inn V\inn\veps_3,p_3\inn N\inn\veps_1)&\sim&4\veps_1^i\eps^{a_0\cdots a_4}(p_2\inn V\inn\veps_3)_{a_4}(\veps_2)_{a_2a_3}(p_2)_{a_0}(p_3)_{i}(p_3)_{a_1}\cI_5\labell{I5}\\
{\cal A}(p_2\inn N\inn\veps_3,p_3\inn N\inn\veps_1)&\sim&4\veps_1^i\eps^{a_0\cdots a_4}(p_2\inn N\inn\veps_3)_{a_4}(\veps_2)_{a_2a_3}(p_2)_{a_0}(p_3)_{i}(p_3)_{a_1}\cI_6\nonumber
\eeqa
where
\beqa
\cI_5
&=&\int d^2z_1d^2z_2d^2z_3\frac{z_{1\bar1}^2K}{|z_{21}|^2|z_{31}|^2|z_{2\bar 1}|^2|z_{3\bar 1}|^2}\big[\cP(z_2,z_3)\pm\cP(\bz_2,\bz_3)-\cP(\bz_2,z_3)\mp \cP(z_2,\bz_3)\big]\nonumber\\
\cI_6
&=&\int d^2z_1d^2z_2d^2z_3\frac{z_{1\bar1}^2K}{|z_{21}|^2|z_{31}|^2|z_{2\bar 1}|^2|z_{3\bar 1}|^2}\big[\cP(z_2,z_3)\pm\cP(\bz_2,\bz_3)+\cP(\bz_2,z_3)\pm \cP(z_2,\bz_3)\big]\nonumber
\eeqa
where the first sign is for symmetric and the second sign is for the antisymmetric polarization tensor $\veps_3$. Using the property of $\cP(z_i,z_j)=-(\cP(\bz_i,\bz_j))^*$, one finds that the integrand is pure imaginary for the symmetric case and is real for the antisymmetric case. The integral for the pure imaginary case is zero. By mapping the integral to unit disk and fixing the $SL(2,R)$ symmetry as in \reef{fix} one can show that the real integrals are the real integrals  that appear in amplitudes \reef{I2} and \reef{I3}, \ie
\beqa
\cI_5&=&-2\cI_2,\,\,\,\cI_6\,=\,2\cI_3\,\nonumber
\eeqa
Moreover, the integrand in $\cI_2$ transforms to the integrand in $\cI_3$ under $(z_2,\bz_2,z_3,\bz_3)\rightarrow (z_3,\bz_3,z_2,\bz_2)$, 
which means  
\beqa
\cI_3(p_1,p_2,p_3)&=&\cI_2(p_1,p_3,p_2)\labell{sym}
\eeqa

The RR tensor in \reef{I5} can be written in terms of $F_{a_0i}$, so they are invariant under the RR gauge transformation. The $\veps_2$ can also be written in terms of the field strength. So they satisfy the Ward identity associated with $\veps_2$. 

The  6 contractions  in $b_{10}$ in which  two momenta contract with each other are
\beqa
{\cal A}_{10}(pp)&\sim&2\frac{6!}{5!}\veps_1^i\eps^{a_0\cdots a_4}\int d^2z_1d^2z_2d^2z_3\frac{z_{1\bar1}^2K}{|z_{21}|^2|z_{31}|^2|z_{2\bar 1}|^2|z_{3\bar 1}|^2}\nonumber\\
&& \bigg(-\cP(z_2,z_3) A_{7[ia_0\cdots a_4]} +\cP(z_2,\bar{z}_2)A_{8[ia_0\cdots a_4]}  
+\cP(\bz_2,{z}_3)  A_{9[ia_0\cdots a_4]}\nonumber\\ 
&&-\cP(z_2,\bar{z}_3) A_{10[ia_0\cdots a_4]} 
+\cP(z_3,{\bz}_3)A_{11[ia_0\cdots a_4]}  -\cP(\bar{z}_2,\bar{z}_3)A_{12[ia_0\cdots a_4]}  \bigg) \labell{amp13}
\eeqa
where
\beqa
A_{7[ia_0\cdots a_4]}&=&p_2\inn p_3(p_2\inn D)_{[i}(p_3\inn D)_{a_0}(\veps_2\inn D)_{a_1a_2}(\veps_3\inn D)_{a_3a_4]}\nonumber\\
A_{8[ia_0\cdots a_4]}&=&p_2\inn D\inn p_2(p_3)_{[i}(p_3\inn D)_{a_0}(\veps_2\inn D)_{a_1a_2}(\veps_3\inn D)_{a_3a_4]}\nonumber\\
A_{9[ia_0\cdots a_4]}&=&p_2\inn D\inn p_3(p_2)_{[i}(p_3\inn D)_{a_0}(\veps_2\inn D)_{a_1a_2}(\veps_3\inn D)_{a_3a_4]}\nonumber\\
A_{10[ia_0\cdots a_4]}&=&p_2\inn D\inn p_3(p_3)_{[i}(p_2\inn D)_{a_0}(\veps_2\inn D)_{a_1a_2}(\veps_3\inn D)_{a_3a_4]}\nonumber\\
A_{11[ia_0\cdots a_4]}&=&p_3\inn D\inn p_3(p_2)_{[i}(p_2\inn D)_{a_0}(\veps_2\inn D)_{a_1a_2}(\veps_3\inn D)_{a_3a_4]}\nonumber\\
A_{12[ia_0\cdots a_4]}&=&p_2\inn p_3(p_2)_{[i}(p_3)_{a_0}(\veps_2\inn D)_{a_1a_2}(\veps_3\inn D)_{a_3a_4]}\nonumber
\eeqa
It is easy to verifies that the above amplitude is zero when both NSNS polarization tensors are symmetric. When one of them is symmetric and the other one is antisymmetric, one finds some non-zero terms. However, the integrand for those terms are pure imaginary which are  zero after integration. When both the NSNS polarization tensors are antisymmetric the above factors simplify to
\beqa
A_{7[ia_0\cdots a_4]}&=&-\frac{p_2\inn p_3}{6}\big[(p_2)_{i}(p_3)_{a_0}-(p_2)_{a_0}(p_3)_{i}\big](\veps_2)_{a_1a_2}(\veps_3)_{a_3a_4}\nonumber\\
A_{8[ia_0\cdots a_4]}&=&\frac{p_2\inn D\inn p_2}{3}(p_3)_{i}(p_3)_{a_0}(\veps_2)_{a_1a_2}(\veps_3)_{a_3a_4}\nonumber\\
A_{9[ia_0\cdots a_4]}&=&\frac{p_2\inn D\inn p_3}{6}\big[(p_2)_{i}(p_3)_{a_0}+(p_2)_{a_0}(p_3)_i\big](\veps_2)_{a_1a_2}(\veps_3)_{a_3a_4}\nonumber\\
A_{10[ia_0\cdots a_4]}&=&\frac{p_2\inn D\inn p_3}{6}\big[(p_2)_{i}(p_3)_{a_0}+(p_2)_{a_0}(p_3)_i\big](\veps_2)_{a_1a_2}(\veps_3)_{a_3a_4}\nonumber\\
A_{11[ia_0\cdots a_4]}&=&\frac{p_3\inn D\inn p_3}{3}(p_2)_{i}(p_2)_{a_0}(\veps_2)_{a_1a_2}(\veps_3)_{a_3a_4}\nonumber\\
A_{12[ia_0\cdots a_4]}&=&\frac{p_2\inn p_3}{6}\big[(p_2)_{i}(p_3)_{a_0}-(p_2)_{a_0}(p_3)_{i}\big](\veps_2)_{a_1a_2}(\veps_3)_{a_3a_4}\nonumber
\eeqa
Replacing them in \reef{amp13}, one finds the following structures:
\beqa 
{\cal A}(p_2p_3)&\sim&2\veps_1^i\eps^{a_0\cdots a_4}\big[(p_2\inn V\inn p_3)(p_2)_{i}(p_3)_{a_0}-(p_2\inn N\inn p_3)(p_2)_{a_0}(p_3)_{i}\big](\veps_2)_{a_1a_2}(\veps_3)_{a_3a_4}\cI_6\nonumber\\
&&+2\veps_1^i\eps^{a_0\cdots a_4}\big[(p_2\inn N\inn p_3)(p_2)_{i}(p_3)_{a_0}-(p_2\inn V\inn p_3)(p_2)_{a_0}(p_3)_{i}\big](\veps_2)_{a_1a_2}(\veps_3)_{a_3a_4}\cI_5\nonumber\\
{\cal A}(p_2\inn D\inn p_2)&\sim&4(p_2\inn D\inn p_2)\veps_1^i\eps^{a_0\cdots a_4}(p_3)_{i}(p_3)_{a_0}(\veps_2)_{a_1a_2}(\veps_3)_{a_3a_4}\cI_7\nonumber\\
{\cal A}(p_3\inn D\inn p_3)&\sim&4(p_3\inn D\inn p_3)\veps_1^i\eps^{a_0\cdots a_4}(p_2)_{i}(p_2)_{a_0}(\veps_2)_{a_1a_2}(\veps_3)_{a_3a_4}\cI_4\labell{I7}
\eeqa
where 
\beqa
\cI_7
&=&\int d^2z_1d^2z_2d^2z_3\frac{z_{1\bar1}^2\cP(z_2,\bz_2)K}{|z_{21}|^2|z_{31}|^2|z_{2\bar 1}|^2|z_{3\bar 1}|^2}\nonumber
\eeqa
Comparing the integrand in this integral with the integrand in $\cI_4$ in \reef{I44}, one finds the relation 
\beqa
\cI_7(p_1,p_2,p_3)=\cI_4(p_1,p_3,p_2)
\eeqa


We have seen that the amplitudes in equations  \reef{I1}, \reef{I2}, \reef{I3}, \reef{I4} and \reef{I5} are invariant under the RR and $\veps_2$ gauge transformations. Hence, the amplitudes in \reef{I7} must also be invariant under these transformations. The terms which have $(p_2)_{a_0}$ are obviously invariant under the $\veps_2$ gauge transformation. The terms which have $(p_3)_{a_0}$  are invariant under the $\veps_3$ gauge transformation which should be considered in the $(2\leftrightarrow 3)$ part of the amplitude \reef{A}. To check that  the above amplitudes  are invariant under the RR  gauge transformation we have to know the relations between the integrals.  Likewise, the combination of the above amplitudes and the amplitudes in \reef{I1}, \reef{I2}, \reef{I3}, \reef{I4} and \reef{I5} must be invariant under  the $\veps_3$ gauge transformation provided that there is a  relation between the integrals. We find these relations by demanding that the amplitude must be invariant under these transformations and then verify them by explicit calculation of the integrals for the restricted case of \reef{kin}. The total amplitude is
\beqa
{\cal A}&\!\!\!\!\!\sim\!\!\!\!\!&8(\veps_1)_i\eps_{a_0\cdots a_4}\bigg(-2(p_1\inn N\inn\veps_3)^{a_3}p_2^{i}p_3^{a_0}\cI_1+4(p_3\inn V\inn\veps_3)^{a_3}p_2^{i}p_3^{a_0}\cI_4+(p_3\inn V\inn p_3)p_2^{i}\veps_3^{a_3a_0}\cI_4
\nonumber\\
&&+(p_2\inn N\inn\veps_3)^{a_3}p_2^{i}p_3^{a_0}\cI_2
+(p_2\inn V\inn\veps_3)^{a_3}p_3^{i}p_3^{a_0}\cI_2+\frac{1}{2}(p_2\inn V\inn p_3)p_3^{i}\veps_3^{a_3a_0}\cI_2\labell{total}\\
&&-(p_2\inn V\inn\veps_3)^{a_3}p_2^{i}p_3^{a_0}\cI_3-(p_2\inn N\inn\veps_3)^{a_3}p_3^{i}p_3^{a_0}\cI_3-\frac{1}{2}(p_2\inn N\inn p_3)p_3^{i}\veps_3^{a_3a_0}\cI_3
\bigg)p_2^{a_2}\veps_2^{a_1a_4}+(2\leftrightarrow 3)\nonumber
\eeqa  
In the special case that the Mandelstam variables are  constrained as \reef{kin}, the above amplitude  satisfies the Ward identity corresponding to $\veps_3$ provided that there is the following relation between $\cI_1$ and $\cI_4$:
\beqa
p_3\inn V\inn p_3\cI_4&=&p_1\inn N\inn p_3\cI_1\labell{iden1}
\eeqa
Similar relation in the $(2\leftrightarrow 3)$ part, \ie $p_2\inn V\inn p_2\cI_7=p_1\inn N\inn p_2\cI_1$. Using these  relations, one observes  that the  terms in \reef{I7}   are invariant under the RR gauge transformation: The first two lines in \reef{I7} are zero under the constraint \reef{kin}, and using conservation of momentum along the brane, the other terms  can be write  as
\beqa
&& {\cal A}(p_2\inn D\inn p_2)+{\cal A}(p_3\inn D\inn p_3)\sim-4(p_2\inn D\inn p_2)\veps_1^i\eps^{a_0\cdots a_4}(p_3)_{i}(p_1)_{a_0}(\veps_2)_{a_1a_2}(\veps_3)_{a_3a_4}\cI_7\nonumber\\
&&-8\big[(p_1\inn N \inn p_2)(\veps_1\inn N \inn p_3)-(p_1\inn N \inn p_3)(\veps_1\inn N \inn p_2)\big]\eps^{a_0\cdots a_4}(p_2)_{a_0}(\veps_2)_{a_1a_2}(\veps_3)_{a_3a_4}\cI_1\nonumber
\eeqa
The first term can be written in terms of field strength $F_{ia_0}$ and the terms in the second line become zero upon replacing the RR polarization with its momentum.

In the general case,
the last two terms in the second and the third lines of \reef{total} each satisfies the Ward identity. The sum of the other  terms then must satisfy the Ward identity. This happens if there are the following relations between the integrals:
\beqa
-2p_1\inn N\inn p_3\cI_1+2p_3\inn V\inn p_3\cI_4+p_2\inn N\inn p_3\cI_2-p_2\inn V\inn p_3\cI_3&=&0\labell{iden2}
\eeqa
From the $(2\leftrightarrow 3)$ part, one finds the following relation:
\beqa
-2p_1\inn N\inn p_2\cI_1+2p_2\inn V\inn p_2\cI_7+p_2\inn N\inn p_3\cI_3-p_2\inn V\inn p_3\cI_2&=&0\labell{iden3}
\eeqa
Using the above two relations, one can show that the terms in \reef{I7} satisfy the Ward identity corresponding to the RR gauge transformation.

One can  use the identities \reef{iden2} and \reef{iden3} to write the amplitude \reef{total} in terms of RR field strength as
\beqa
{\cal A}&\!\!\!\!\!\sim\!\!\!\!\!&8\eps_{a_0\cdots a_4}(F_1)_{a_0i}\bigg(-2(p_1\inn N\inn\veps_3)^{a_3}p_2^{i}p_3^{a_2}\cI_1+4(p_3\inn V\inn\veps_3)^{a_3}p_2^{i}p_3^{a_2}\cI_4+(p_2\inn N\inn\veps_3)^{a_3}p_2^{i}p_3^{a_2}\cI_2
\nonumber\\
&&
+(p_2\inn V\inn\veps_3)^{a_3}p_3^{i}p_3^{a_2}\cI_2
-(p_2\inn V\inn\veps_3)^{a_3}p_2^{i}p_3^{a_2}\cI_3-(p_2\inn N\inn\veps_3)^{a_3}p_3^{i}p_3^{a_2}\cI_3\bigg)\veps_2^{a_1a_4}+(2\leftrightarrow 3)\nonumber\\
&&-8\eps_{a_0\cdots a_4}(F_1)_{a_0i}\bigg(p_3\inn V\inn p_3\,p_2^{i}\veps_3^{a_3a_2}\cI_4+\frac{1}{2}p_2\inn V\inn p_3\,p_3^{i}\veps_3^{a_3a_2}\cI_2-\frac{1}{2}p_2\inn N\inn p_3\,p_3^{i}\veps_3^{a_3a_2}\cI_3
\bigg)\veps_2^{a_1a_4}\nonumber\\
&&+8\eps_{a_0\cdots a_4}(F_1)_{ij}p_2^ip_3^j\veps_2^{a_3a_0}\veps_3^{a_1a_4}p_3^{a_2}\cI_1\labell{totalfF}
\eeqa  
As a double check, we have calculated the amplitude \reef{amp2} in the $(-1/2,-1/2)$-picture  and  found exact agreement with the above result.

Using the relation \reef{iden2} and the following identity:
\beqa
\eps_{a_0\cdots a_4} p_{\mu}H_3^{\mu a_3a_0}&=&\eps_{a_0\cdots a_4}(2p_{\mu}\veps_3^{\mu a_3}p_3^{a_0}+(p_{\mu} p_3^{\mu})\veps_3^{a_3a_0})\nonumber
 \eeqa
 where $p_{\mu}$ is any vector, one can write the amplitude \reef{total} in terms of field strength $H_2,H_3$,
\beqa
{\cal A}&\sim&\frac{4}{3}\eps_{a_0\cdots a_4}(\veps_1)_i\bigg(\big[2(p_1\inn N\inn H_3)^{a_3a_0}p_2^{i}\cI+4(p_3\inn V\inn H_3)^{a_3a_0}p_2^{i}\cI_4
\nonumber\\
&&\qquad\qquad\qquad+(p_2\inn N\inn H_3)^{a_3a_0}p_2^{i}\cI_2
+(p_2\inn V\inn H_3)^{a_3a_0}p_3^{i}\cI_2\big]H_2^{a_1a_4a_2}\labell{totalf}\\
&&\qquad\qquad\qquad-\big[(p_3\inn N\inn H_2)^{a_3a_0}p_2^{i}\cI_2+(p_3\inn V\inn H_2)^{a_3a_0}p_3^{i}\cI_2\big]H_3^{a_1a_4a_2}
\bigg)+(2\leftrightarrow 3)\nonumber
\eeqa 
where we have also used the relations \reef{iden0} and \reef{sym}.

Mapping the integrand in each of the integrals $\cI_2,\, \, \cI_4$  to unit disk and fixing the $SL(2,R)$ symmetry as in \reef{fix}, one finds that the integral over $\theta$ gives zero result when $\cI_2$ is  pure imaginary. The real integrals are
\beqa
\cI_2&=&2\int_0^1dr_2\int_0^{1}dr_3\frac{(1-r_2^2)}{r_2}\int_0^{2\pi}d\theta\frac{[r_3(1+r_2^2)-r_2(1+r_3^2)\cos(\theta)]\tK}{|1-r_2r_3e^{i\theta}|^2|r_2-r_3e^{-i\theta}|^2}\nonumber\\
\cI_4&=&-\int_0^1dr_2\int_0^{1}dr_3\frac{(1+r_3^2)}{r_2r_3(1-r_3^2)}\int_0^{2\pi}d\theta\tK\labell{integ}
\eeqa

Up to this point the result is valid for the general case. However, to be able to perform the integrals we restrict the Mandelstam variables to \reef{kin}. 
This restriction makes the $\theta$-integral to be simple, \ie
\beqa
\cI_2&=&4\pi\int_0^1dr_3\int_0^{r_3}dr_2\frac{\tK}{r_2r_3}\nonumber\\
\cI_4&=&-2\pi\int_0^1dr_2\int_0^{1}dr_3\frac{(1+r_3^2)\tK}{r_2r_3(1-r_3^2)}\labell{integ2}
\eeqa
We have used the Maple to evaluate the $\theta$-integral in $\cI_2$. Using the definition of the beta function, the radial integrals in $\cI_4$ gives two beta functions, \ie
\beqa
\cI_4&=&-\pi \frac{p_1\inn N\inn p_3}{p_3\inn D\inn p_3}B(p_1\inn p_2, 1+p_2\inn D\inn p_2)B(p_1\inn p_3, 1+p_3\inn D\inn p_3)\labell{integ147}
\eeqa
Using the relations \reef{amp51} and \reef{iden0}, one observes that $\cI_1,\,\cI_4$  satisfy the relation \reef{iden1} as expected. 

Using  the formula in the Appendix, one finds the radial integrals in $\cI_2$ to be 
\beqa
\cI_2&=&\pi\frac{B(p_1\inn p_2+p_1\inn p_3,1+p_3\inn D\inn p_3)}{p_1\inn p_2}{}_3F_2\bigg[{p_1\inn p_2+p_1\inn p_3,\ p_1\inn p_2, \ -p_2\inn D\inn p_2 \atop 1+p_1\inn p_2+p_1\inn p_3+p_3\inn D\inn p_3, \ 1+p_1\inn p_2}\ ;\ 1\bigg]\nonumber
\eeqa

 To study the low energy limit of the amplitude \reef{totalf}, we expand $\cI_2,\cI_4$  at the low energy. The expansion of beta function is standard and for expanding the hypergeometric function we use the package \cite{Huber:2005yg}. The result is 
\beqa
\cI_2&\!\!\!\!\!=\!\!\!\!\!&\pi\left(\frac{1}{p_1\inn p_2(p_1\inn p_2+p_1\inn p_3)}-\frac{\pi^2}{6}\frac{p_3\inn D\inn p_3}{p_1\inn p_2}+\cdots\right)\nonumber\\
\cI_4&\!\!\!\!\!=\!\!\!\!\!&-\pi \left(\frac{1}{2p_1\inn p_2 p_1\inn p_3}+\frac{1}{p_1\inn p_2\,p_3\inn D\inn p_3}-\frac{\pi^2}{12}\bigg[\frac{2p_1\inn N\inn p_3}{p_1\inn p_2}+\frac{2p_2\inn D\inn p_2}{p_3\inn D\inn p_3}+\frac{p_2\inn D\inn p_2}{p_1\inn p_3}\bigg]+\cdots\right)\nonumber
\eeqa
 From these expansions and the expansion \reef{amp6} for $\cI$, one realizes that only $\cI_4$ has massless open string pole at order $O(\alpha'^2)$. Hence, the amplitude \reef{totalf} has the following massless open string pole at order $O(\alpha'^2)$:
 \beqa
 {\cal A}^f&\sim&\frac{8\pi^3}{9}\frac{p_2\inn D\inn p_2}{p_3\inn D\inn p_3}\eps_{a_0\cdots a_4}(\veps_1)_i(p_3\inn V\inn H_3)^{a_3a_0}p_2^{i}H_2^{a_1a_4a_2}
+(2\leftrightarrow 3)\nonumber
\eeqa 
This is exactly the same as the amplitude \reef{ampfield} in field theory side. This ends our illustration of the consistency of the couplings in \reef{first111} with the string theory S-matrix element of one RR and two NSNS states.

 {\bf Acknowledgments}:  This work is supported by Ferdowsi University of Mashhad.   

\newpage
{\bf\Large {Appendix}}

In this appendix we calculate the following integral:
\beqa
I&=&\int_0^1dx\int_0^xdy\,x^ay^b(1-x)^c(1-y)^d
\eeqa
To take the integral, one should first change the variable to $y=xu$, \ie
\beqa
I&=&\int_0^1dx\,x^{1+a+b}(1-x)^c\int_0^1du\, u^b(1-xu)^d
\eeqa
 Then using the integral representation of the generalized hypergeometric function ${}_pF_q$ \cite{ISG}:
\beqa
{}_2F_1\bigg[{-c,\ 1+a,\atop  2+a+b}\ ;\ x 
\bigg]B(1+a,1+b)&\!\!\!\!\!=\!\!\!\!\!&\int_0^1du\,u^a(1-u)^b(1-xu)^c\nonumber\\
{}_{p+1}F_{q+1}\bigg[{1+a,\ a_1, \ \cdots, a_p\atop  2+a+b,\ b_1, \ \cdots, \ b_q}\ ;\ \la 
\bigg]B(1+a,1+b)&\!\!\!\!\!=\!\!\!\!\!&\int_0^1dx\,x^a(1-x)^b{}_{p}F_{q}\bigg[{ a_1, \ \cdots, a_p\atop   b_1, \ \cdots, \ b_q}\ ;\ \la 
\bigg]\nonumber
\eeqa
one finds
\beqa
I&=&\frac{B(1+c,2+a+b)}{1+b}{}_3F_2\bigg[{2+a+b,\ 1+b, \ -d \atop 3+a+b+c, \ 2+b}\ ;\ 1\bigg]
\eeqa


\end{document}